\documentclass[10pt,a4paper,titlepage]{article}
\usepackage{latex8}
\usepackage{times}
\usepackage{epsfig, graphicx, subfigure}
\usepackage{color}
\usepackage{psfrag}
\usepackage{amsmath, amsthm, amssymb}
\usepackage{latexsym}
\usepackage[T1]{fontenc}
\usepackage[english]{babel}
\usepackage{fancyheadings}
\usepackage{algorithm}
\usepackage{algorithmic}

\usepackage{a4wide}
\newcommand{\remove}[1]{}

\newtheorem{theorem}{\bf Theorem}
\newtheorem{lemma}{\bf Lemma}

\newtheorem{definition}{\bf Definition}

\renewcommand{\algorithmicrequire}{\textbf{Input:}}
\renewcommand{\algorithmicensure}{\textbf{Output:}}
\renewcommand{\algorithmiccomment}[1]{// #1}

\newcounter{num}

\newcommand{\squishlist}{
   \begin{list}{$\bullet$}
    { \setlength{\itemsep}{0pt}      \setlength{\parsep}{3pt}
      \setlength{\topsep}{3pt}       \setlength{\partopsep}{0pt}
      \setlength{\leftmargin}{1.5em} \setlength{\labelwidth}{1em}
      \setlength{\labelsep}{0.5em} } }

\newcommand{\squishlisttwo}{
   \begin{list}{$\bullet$}
    { \setlength{\itemsep}{0pt}    \setlength{\parsep}{0pt}
      \setlength{\topsep}{0pt}     \setlength{\partopsep}{0pt}
      \setlength{\leftmargin}{2em} \setlength{\labelwidth}{1.5em}
      \setlength{\labelsep}{0.5em} } }

\newcommand{\squishend}{
    \end{list}  }

\begin{document}
\onecolumn

\begin{titlepage}
\title{
  \raisebox{30mm}[0mm][0mm]{\Large
    Technical Report no. 2008-69
  }
  \raisebox{5mm}[0mm][0mm]{
    \textbf{
      \begin{tabular}{c}
      NB-FEB: An Easy-to-Use and Scalable Universal \\
		Synchronization Primitive for Parallel Programming
      \end{tabular}
    }
  }
}
\author{\raisebox{-15mm}[0mm][0mm]{\textbf{\Large Phuong Hoai Ha}} \and \raisebox{-15mm}[0mm][0mm]{\textbf{\Large Philippas Tsigas\footnote{Department of Computer Science and Engineering, Chalmers University of Technology, SE-412 96 G{\"o}teborg, Sweden.}}} \and \raisebox{-15mm}[0mm][0mm]{\textbf{\Large Otto J. Anshus}}}
\date{
  \vspace{\stretch{1}}
  \enlargethispage{1.1\baselineskip}
  {\resizebox*{0.2\columnwidth}{!}{\includegraphics[angle=90]{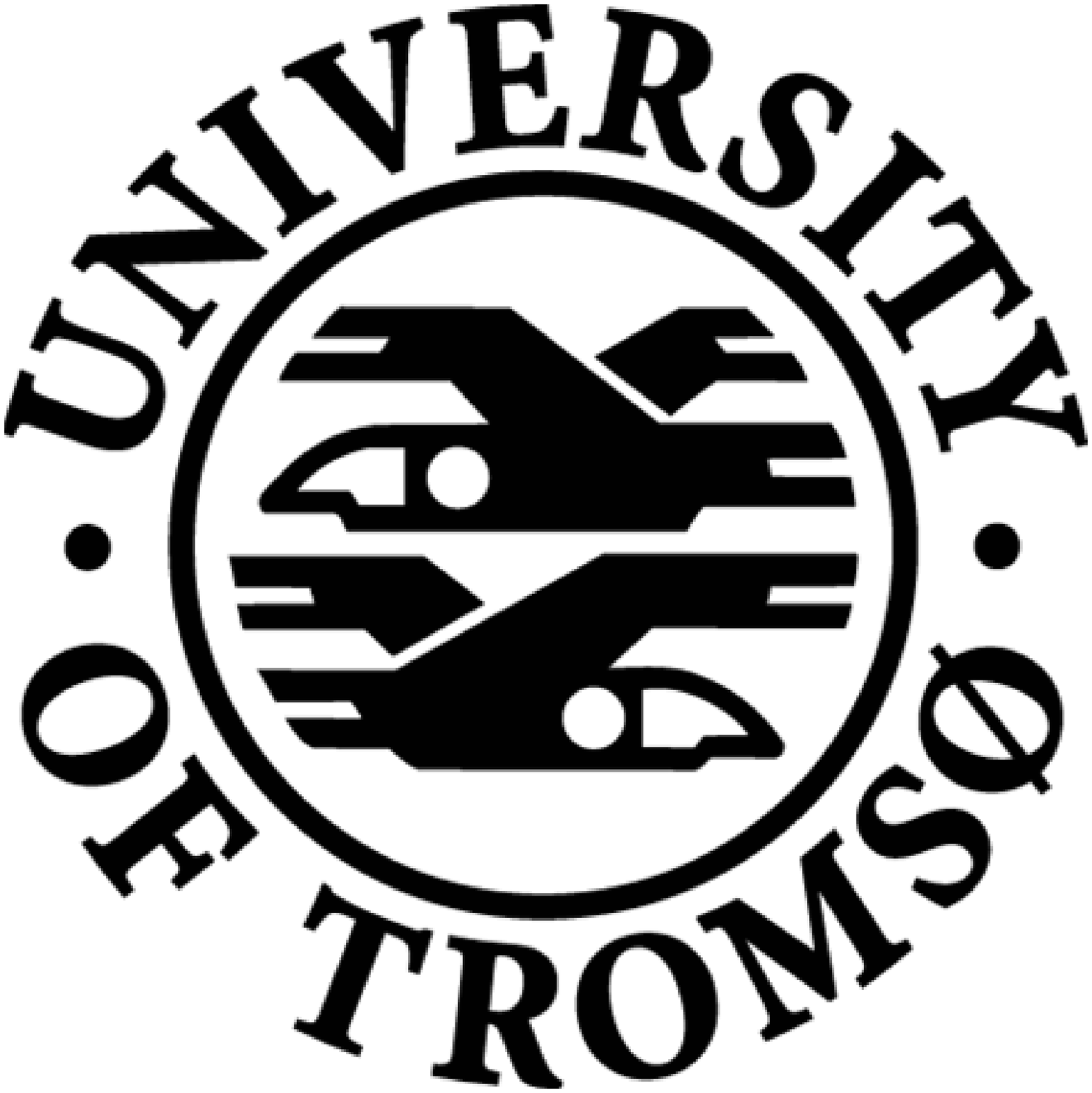}}} \\
  \vspace{12mm}
  Department of Computing Science \\
  Faculty of Science \\
  University of Troms{\o} \\
  N-9037 Troms{\o}, Norway \\
  \vspace{12mm}
   Troms{\o}, October 2008.
}    
\maketitle
\end{titlepage}

\newpage
\thispagestyle{empty}
\mbox{}
\vspace{\stretch{1}}

\noindent
{\large
  \begin{tabular}{l}
    Technical Report in Computing Science at \\
    University of Troms{\o}
    \vspace{3ex} \\
    Technical Report no. 2008-69 \\
    ISSN: XXXX-XXXX 
    \vspace{3ex} \\
    Department of Computing Science \\
    Faculty of Science \\
    University of Troms{\o} \\
    N-9037 Troms{\o}, Norway \\
    \vspace{3ex} \\
    Troms{\o}, Norway, October 2008.
  \end{tabular}
}

\newpage

\twocolumn

\begin{abstract}
This paper addresses the problem of universal synchronization primitives that can support {\em scalable} thread synchronization for large-scale many-core architectures.
The universal synchronization primitives that have been deployed widely in conventional architectures, are the {\em compare-and-swap} (CAS) and {\em load-linked/store-conditional} (LL/SC) primitives. However, such synchronization primitives are expected to reach their scalability limits in the evolution to many-core architectures with thousands of cores.

We introduce a {\em non-blocking} full/empty bit primitive, or NB-FEB for short, as a promising synchronization primitive for parallel programming on may-core architectures. We show that the NB-FEB primitive is {\em universal, scalable, feasible} and {\em convenient} to use. NB-FEB, together with registers, can solve the consensus problem for an arbitrary number of processes ({\em universality}). NB-FEB is {\em combinable}, namely its memory requests to the same memory location can be combined into only one memory request, which consequently mitigates performance degradation due to synchronization "hot spots" ({\em scalability}). Since NB-FEB is a variant of the original full/empty bit that always returns a value instead of waiting for a conditional flag, it is as feasible as the original full/empty bit, which has been implemented in many computer systems ({\em feasibility}). 
The original full/empty bit is well-known as a {\em special-purpose} primitive for fast producer-consumer synchronization and has been used extensively in the specific domain of applications. 
In this paper, we show that NB-FEB can be deployed easily as a {\em general-purpose} primitive. Using NB-FEB, we construct a non-blocking software transactional memory system called NBFEB-STM, which can be used to handle concurrent threads {\em conveniently}. NBFEB-STM is space efficient: the space complexity of each object updated by $N$ concurrent threads/transactions is $\Theta(N)$, the optimal.


\end{abstract}




\section{Introduction}

Universal synchronization primitives \cite{Her91} are essential for constructing non-blocking synchronization mechanisms for parallel programming, like non-blocking software transactional memory \cite{FraH07, HarF03, HerLMS03,MarSS05,RieFF06}.  Non-blocking synchronization eliminates the concurrency control problems of mutual exclusion locks, such as priority inversion, deadlock and convoying.
As many-core architectures with thousands of cores are expected to be our future chip architectures \cite{Asa06}, universal synchronization primitives that can support scalable thread synchronization for such large-scale architectures are desired.   


However, the conventional universal primitives like {\em compare-and-swap} ($CAS$) and {\em load-linked/store-conditional} ($LL/SC$) are expected to reach their scalability limits in the evolution to many-core architectures with thousands of cores.
For each shared memory location, the $LL/SC$ implementation conceptually associates a reservation bit with each processor. The reservations are invalidated when the location are modified by any processor. Implementing $LL/SC$ in the memory (without compromising its semantics) limits the scalability of the multiprocessor since the total directory size increases quadratically with the number of processors \cite{MicS95}. Therefore, the $LL/SC$ primitives are built on conventional cache-coherent protocols \cite{MicS95,CulSG98}. 
However, experimental studies have shown that the $LL/SC$ primitives are not scalable for multicore architectures \cite{SriRK07}. The conventional cache-coherent protocols are considered inefficient for large scale manycore architectures \cite{Asa06}. As a result, several emerging multicore architectures like the NVIDIA CUDA \cite{CUDA}, the ClearSpeed CSX \cite{CSX06}, the IBM Cell BE \cite{GscHFHWY06} and the Cyclops-64 \cite{CasCSW02} architectures utilize fast local memory for each processing core rather than coherent data cache. 

For the emerging many-core architectures without coherent data cache, the $CAS$ primitive is not scalable either since $CAS$ is not {\em combinable} \cite{KruRS88,BleGV08}. Primitives are combinable if their memory requests to the same memory location (arriving at a switch of the processor-to-memory interconnection network) can be combined into only one memory request. Separate replies to the original requests are later created from the reply to the combined request (at the switch). The combining technique has been implemented in the NYU Ultracomputer \cite{GotGKMRS82} and the IBM RP3 \cite{PfiBGHKMMNW85} machine and has been shown to be a promising technique for large-scale multiprocessors to alleviate the performance degradation due to synchronization "hot spot". Although the {\em single-valued} $CAS_{a}(x,b)$ \cite{BleGV08}, which will atomically swap $b$ to $x$ if $x$ equals $a$ is combinable, the number of instructions $CAS_{a}$ must be as many as the number of integers $a$ that can be stored in one memory word (e.g. $2^{64}$ $CAS_{a}$ instructions for 64-bit words). This fact makes the {\em single-valued} $CAS_{a}$ unfeasible for hardware implementation.

Another universal primitive called {\em sticky bit} has been suggested in \cite{Plo89}, but it has not been deployed so far due to its usage complexity. To the best of our knowledge, the universal construction using the sticky bit \cite{Plo89} does not prevent a delayed thread, even after being helped, from jamming the sticky bits of a cell that has been re-initialized and reused. 
Since the universal construction is built on a doubly-linked list of cells, it is not obvious how an external garbage collector (supported by the underlying system) can help solve the problem. Moreover, the space complexity of the universal construction for an object is as high as $O(N^2logN)$, where $N$ is the number of processes.

This paper suggests a novel synchronization primitive, called NB-FEB, as a promising synchronization primitive for parallel programming on many-core architectures. What makes NB-FEB be a promising primitive is its following four main properties. NB-FEB is:
\begin{description}
\item[Feasible]: NB-FEB is a {\em non-blocking} variant of the conventional full/empty bit that {\em always returns} the old value of the variable instead of waiting for its conditional flag to be set (or cleared). 
This simple modification makes NB-FEB as {\em feasible} as the original (blocking) full/empty bit, which has been implemented in many computer systems like HEP \cite{Smi85}, Tera \cite{AlvCCKPS90}, MDP \cite{DalFKLNNDF92}, Sparcle \cite{AKKLYDP93}, M-Machine \cite{KecDMCCL98} and Eldorado \cite{FeoHKK05}. The space overhead of full/empty bits can be reduced using the {\em synchronization state buffer (SSB)} \cite{ZhuSHG07}.

\item[Universal]: This simple modification, however, significantly increases the synchronization power of full/empty bits, making NB-FEB as powerful as $CAS$ or $LL/SC$. NB-FEB, together with registers, can solve consensus problem for arbitrary number of processes, the essential property for constructing non-blocking synchronization mechanisms (cf. Section \ref{sec:universality}). 

\item[Scalable]: Like the original full/empty bit, NB-FEB is {\em combinable}: its memory requests to the same memory location can be combined into only one memory request (cf. Section \ref{sec:combinability}). This empowers NB-FEB 
with the ability to provide {\em scalable} thread synchronization for large-scale many-core architectures.

\item[Convenient to use]: The original full/empty bit is well-known as a {\em special-purpose} primitive for fast producer-consumer synchronization and has been used extensively in the specific domain of applications. 
In this paper, we show that NB-FEB can be deployed easily as a {\em general-purpose} primitive. Using NB-FEB, we construct a non-blocking software transactional memory system called NBFEB-STM, which can be used to handle concurrent threads {\em conveniently}. NBFEB-STM is space efficient: the space complexity of each object updated by $N$ concurrent threads/transactions is $\Theta(N)$, the optimal (cf. Section \ref{sec:NBFEBSTM}). 
\end{description}




\remove{ 
\begin{algorithm}[tbh]
  \caption{{\sc LSC}($x$: variable): a non-blocking variant of the original Load-if-Set-and-Clear operation, which returns $\perp$ (instead of waiting) if $flag_x$ is false.} \label{alg:LSC}
	\begin{algorithmic}
	\IF{ $flag_x =$ \TRUE}
	\STATE $flag_x \leftarrow$ \FALSE;
	\STATE $tmp \leftarrow x$; \COMMENT{Read the value of $x$}
	\RETURN $tmp$; 
	\ELSE 
	\RETURN $\perp$;
	\ENDIF
  	\end{algorithmic}
\end{algorithm}
} 





The rest of this paper is organized as follows. 
Section \ref{sec:models} presents the shared memory and interconnection network models assumed in this paper. 
Sections \ref{sec:NBFEB} describes the NB-FEB primitive in detail and proves its universality and combinability properties.
Section \ref{sec:NBFEBSTM} presents NBFEB-STM, the obstruction-free multi-versioning STM constructed on the NB-FEB primitive.
Section \ref{sec:GarbageCollector} describes a garbage collector that can be used as an external garbage collector for the NBFEB-STM.


\section{Models} \label{sec:models}

As previous research on the synchronization power of synchronization primitives \cite{Her91}, this paper assumes the linearizable shared memory model \cite{AttW04}. 
Due to NB-FEB combinability, as in \cite{KruRS88} we assume that the processor-to-memory interconnection network is {\em nonovertaking} and that a reply message is sent back on the same path followed by the request message. The immediate nodes, on the communication path from a processor to a global shared memory module (such as switches of a multistage interconnection network or higher memory modules of a multilevel memory hierarchy), can detect requests destined for the same destination and maintain the queues of requests. 
No memory coherent schemes are assumed.






\section{NB-FEB Primitives} \label{sec:NBFEB}

The set of NB-FEB primitives consists of four sub-primitives: $TFAS$ (Algorithm \ref{alg:TFAS}), $Load$ (Algorithm \ref{alg:Load}), $SAC$ (Algorithm \ref{alg:SAC}) and $SAS$ (Algorithm \ref{alg:SAS}). The last three primitives are similar to those of the original full/empty bit. Regarding conditional load primitives, a processor can
check the flag value, $flag_x$, returned by the unconditional load primitive to determine if it was successful.

\begin{algorithm}[t]
  \caption{{\sc TFAS}($x$: variable, $v$: value): Test-Flag-And-Set, a non-blocking variant of the original Store-if-Clear-and-Set primitive, which {\em always} returns the old value of $x$.} \label{alg:TFAS}
	\begin{algorithmic}
	\STATE $(o, flag_o) \leftarrow (x, flag_x)$; 
	\IF{$flag_x =$ \FALSE}
	\STATE $(x, flag_x) \leftarrow (v,$ \TRUE $)$;
	\ENDIF
	\RETURN $(o, flag_o)$; 
  	\end{algorithmic}
\end{algorithm}

\begin{algorithm}[t]
  \caption{{\sc Load}($x$: variable)} \label{alg:Load}
	\begin{algorithmic}
	\RETURN $(x,flag_x)$;
  	\end{algorithmic}
\end{algorithm}

\begin{algorithm}[t!]
  \caption{{\sc SAC}($x$: variable, $v$: value): Store-And-Clear} \label{alg:SAC}
	\begin{algorithmic}
	\STATE $(o,flag_o) \leftarrow (x,flag_x)$;
	\STATE $(x,flag_x) \leftarrow (v,$ \FALSE $)$;
	\RETURN $(o,flag_o)$;
  	\end{algorithmic}
\end{algorithm}

\begin{algorithm}[t!]
  \caption{{\sc SAS}($x$: variable, $v$: value): Store-And-Set} \label{alg:SAS}
	\begin{algorithmic}
	\STATE $(o,flag_o) \leftarrow (x,flag_x)$;
	\STATE $(x,flag_x) \leftarrow (v,$ \TRUE $)$;
	\RETURN $(o,flag_o)$;
  	\end{algorithmic}
\end{algorithm}

When the value of $flag_x$ returned is not needed, we just write $r \leftarrow$ {\sc TFAS}$(x,v)$ instead of $(r,flag_r) \leftarrow$ {\sc TFAS}$(x,v)$, where $r$ is $x$'s old value. The same applies to $SAC$ and $SAS$. For $Load$, we just write $r \leftarrow x$ instead of $r \leftarrow$ {\sc Load}$(x)$. 
In this paper, the flag value returned is needed only for combining NB-FEB primitives. 

\subsection{$TFAS$: A Universal Primitive}\label{sec:universality}

\begin{lemma}
(Universality) The {\em test-flag-and-set} primitive (or $TFAS$ for short) is universal.
\end{lemma}
\begin{proof}
We will show that there is a wait-free\footnote{An implementation is {\em wait-free} if it guarantees that any process can complete any operation on the implemented object in a finite number of steps, regardless of the execution speeds on the other processes \cite{Her91,Lamport77}.} consensus algorithm, for arbitrary number of processes, that uses only the $TFAS$ primitive and registers. 

The wait-free consensus algorithm is shown in Algorithm \ref{alg:TFASConsensus}.
Processes share a variable called $Decision$, which is initialized to $\perp$ with a $false$ flag. Each process $p$ proposes its value ($\neq \perp$) called $proposal$ by calling {\sc TFAS\_Consensus}$(proposal)$. 

The {\sc TFAS\_Consensus} procedure is clearly wait-free since it contains no loops. We need to prove that i) the procedure returns the same value to all processes and ii) the value returned is the value proposed by some process. Indeed, the procedure will return the proposal of the first process executing $TFAS$ on the $Decision$ variable to all processes. Let $p$ be a process calling the procedure.
\begin{itemize}
\item If $p$ is the first process executing $TFAS$ on the $Decision$ variable, since the $Decision$ variable is initialized to $\perp$ with a $false$ flag, 
$p$'s $TFAS$ will successfully write $p$'s proposal to $Decision$ and return $\perp$, the previous value of $Decision$. Since the value returned is $\perp$, the procedure returns $p$'s proposal (line \ref{alg:T:proposal}T), the proposal of the first process executing $TFAS$.
\item If $p$ is not the first process executing $TFAS$ on the $Decision$ variable, $p$'s $TFAS$ will fail to write $p$'s proposal to $Decision$ since $flag_{Decision}$ has been set to $true$ by the first $TFAS$ on $Decision$. $p$'s $TFAS$ will return the value, called $first$, written by the first $TFAS$. The $first$ value is the proposal of the first process executing $TFAS$ on the $Decision$ variable. Since $first \neq \perp$ (due to the hypothesis that proposals are not $\perp$), the procedure will return $first$ (line \ref{alg:T:first}T).
\end{itemize}
\end{proof}

\begin{algorithm}[t]
  \caption{{\sc TFAS\_Consensus}($proposal$: value)} \label{alg:TFASConsensus}
	$Decision$: shared variable. The shared variable is initialized to $\perp$ with a clear flag (i.e. $flag_{Decision} =$ {\bf false}).\\

  	\algsetup{linenodelimiter=T:}
  	\begin{algorithmic}[1]
	\ENSURE a value agreed by all processes.
	\STATE $first \leftarrow$ {\sc TFAS}$(Decision, proposal)$;
	\IF{ $first = \perp$}
		\RETURN $proposal$; \label{alg:T:proposal}
	\ELSE
		\RETURN $first$; \label{alg:T:first}
	\ENDIF 
  	\end{algorithmic}
\end{algorithm}

\subsection{Combinability}\label{sec:combinability}

\begin{lemma}
(Combinability) NB-FEB primitives are combinable.
\end{lemma}
\begin{proof}
Table \ref{tab:Combination} summarizes the combining logic of NB-FEB primitives on a memory location $x$. The first column is the name of the first primitive request and the first row is the name of  the successive primitive request. For instance, the cell $[SAS,TFAS]$ is the combining logic of $SAS$ and $TFAS$ in which $SAS$ is followed by $TFAS$.
Let $v_1, v_2, r$ and $f_r$ be the value of the first primitive request, the value of the second primitive request, the value returned and the flag returned, respectively. 
In each cell, the first line is the combined request, the second is the reply
to the first primitive request and the third (and forth) is the reply
to the successive primitive request. The values $0$ and $1$ of $f_r$ in the reply represent $false$ and $true$, respectively.

Consider the cell $[TFAS,TFAS]$ as an example. The cell describes the case where request $TFAS(x,v_1)$  is followed by request $TFAS(x,v_2)$, at a switch of the processor-to-memory interconnection network. The two requests can be combined into only one request $TFAS(x,v_1)$ (line 1), which will be forwarded further to the corresponding memory controller. When receiving
 a reply $(r,f_r)$ to the combined request, the switch at which the requests were combined, creates separate replies to the two original requests. The reply to the first original request, $TFAS(x,v_1)$, is $(r,f_r)$ (line 2) as if the request was executed by the memory controller. The reply to the successive request, $TFAS(x,v_2)$, depends on whether the combined request $TFAS(x,v_1)$ has successfully updated the memory location $x$. 
If $f_r=0$, $TFAS(x,v_1)$ has successfully updated $x$ with its value $v_1$. Therefore, the reply to the successive request $TFAS(x,v_2)$ is $(v_1,1)$ as if the request was executed right after the first request $TFAS(x,v_1)$.
If $f_r=1$, $TFAS(x,v_1)$ has failed to update the $x$ variable. Therefore, the reply to the successive request $TFAS(x,v_2)$ is $(r,1)$.


\end{proof}

\begin{figure}[t]
\centering
\begin{tabular}{|l|l|l|l|l|} \hline
 $(x,[v_1])$ & \multicolumn{4}{c|}{ The successive primitive with parameters $(x,[v_2])$}\\
			& $Load$ 		& 	$SAC$ 		& $SAS$ 			& $TFAS$ \\ \hline
$Load$ 	&  $Load$ 		& $SAC(v_2)$ 	& $SAC(v_2)$	& $TFAS(v_2)$\\ 
			& $(r,f_r)$ 	& $(r,f_r)$ 	& $(r,f_r)$		& $(r,f_r)$\\
			& $(r,f_r)$ 	& $(r,f_r)$ 	& $(r,f_r)$		& $(r,f_r)$ \\ \hline
$SAC$ 	& $SAC(v_1)$ 	& $SAC(v_2)$ 	& $SAS(v_2)$ 	& $SAS(v_2)$\\
			& $(r,f_r)$ 	& $(r,f_r)$ 	& $(r,f_r)$		& $(r,f_r)$\\
			& $(v_1,0)$ 	& $(v_1,0)$ 	& $(v_1,0)$		& $(v_1,0)$ \\ \hline
$SAS$ 	& $SAS(v_1)$ 	& $SAC(v_2)$ 	& $SAS(v_2)$ 	& $SAS(v_1)$\\
			& $(r,f_r)$ 	& $(r,f_r)$ 	& $(r,f_r)$		& $(r,f_r)$\\
			& $(v_1,1)$ 	& $(v_1,1)$ 	& $(v_1,1)$		& $(v_1,1)$ \\ \hline
$TFAS$ 	& $TFAS(v_1)$ 	& $SAC(v_2)$ 	& $SAS(v_2)$ 	& $TFAS(v_1)$\\
			& $(r,f_r)$ 	& $(r,f_r)$ 	& $(r,f_r)$		& $(r,f_r)$\\
			& Like 5th 		& Like 5th		& Like 5th		& if $f_r$=0: $(v_1,1)$ \\
			& column 		& column			& column  		& else: $(r,1)$ \\ \hline
\end{tabular}
\caption{The combining logic of NB-FEB primitives on a memory location $x$}\label{tab:Combination}
\end{figure}

\section{NBFEB-STM: Obstruction-free Multi-versioning STM} \label{sec:NBFEBSTM}

Like previous obstruction-free multi-versioning STM called LSA-STM \cite{RieFF06}, the new software transactional memory called NBFEB-STM, assumes that objects are only accessed and modified within transactions. NBFEB-STM assumes that there are no nested transactions, namely each thread executes only one transaction at a time. 
NBFEB-STM, like other obstruction-free STMs \cite{HerLMS03, MarSS05, RieFF06}, is designed for garbage-collected programming languages (e.g. Java).
A variable reclaimed by the garbage collector is assumed to have all bits 0 when it is reused. Note that there are non-blocking garbage collection algorithms that do not require synchronization primitives other than reads and writes while they still guarantee the non-blocking property for application-threads. Such a garbage collection algorithm is presented in Section \ref{sec:GarbageCollector}.

Only two NB-FEB primitives, $TFAS$ and $SAC$, are needed for implementing NBFEB-STM.

\subsection{Challenges and Key Ideas}

Unlike the STMs using $CAS$ \cite{HerLMS03, MarSS05, RieFF06}, NBFEB-STM using $TFAS$ and $SAC$ must handle the problem that $SAC$'s interference with  concurrent $TFAS$es will violate the atomicity semantics expected on variable $x$. Overlapping $TFAS_1$ and $TFAS_2$ both may successfully write their new values to $x$ if $SAC$ interference occurs.

\begin{figure}[t]
\begin{center}
    {\scalebox{0.45}{\input{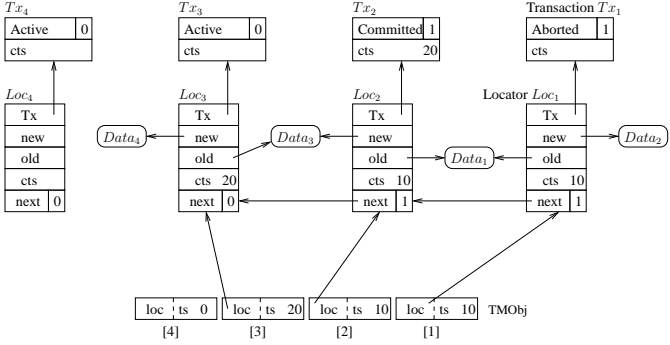}}
    \caption{The data structure of a transactional memory object $TMObj$ in NBFEB-STM with four threads.} \label{fig:NBFEB-STM}}
\end{center}
\end{figure}

The key idea is not to use the transactional memory object $TMObj$ \cite{HerLMS03, MarSS05, RieFF06} that needs to switch its pointer frequently to a new locator (when a transaction commits). Such a $TMObj$ would need $SAC$ in order to clear the pointer's flag, allowing the next transaction to switch the pointer. Instead, NBFEB-STM keeps a linked-list of locators for each object and integrates a write-once pointer $next$ into each locator (cf. Figure\ref{fig:NBFEB-STM}). When opening an object $O$ for write, a transaction $T$ tries to append its locator to $O$'s locator-list by changing the $next$ pointer of the head-locator of the list using $TFAS$. Due to the semantics of $TFAS$, only one of the concurrent transactions trying to append their locators succeeds. The other transactions must retry in order to find the new head and then append their locators to the new head. Using the locator-list, each $next$ pointer is changed only once and thus its flag does not need to be cleared during the lifetime of the corresponding locator. This prevents a $SAC$ from interleaving with concurrent $TFAS$es.
The $next$ pointer, together with its locator, will be reclaimed by the garbage collector when the  lifetime of its locator is over. The garbage collector ensures that a locator will not be recycled until no thread/transaction has a reference to it.

Linking locators together creates another challenge on the space complexity of NBFEB-STM. Unlike the STMs using $CAS$, a delayed/halted transaction $T$ in NBFEB-STM may prevent all locators appended after its locator in a locator-list from being reclaimed. As a result, $T$ may make the system run out of memory and thus prevent other transactions from making progress, violating the obstruction-freedom property.
The key idea to solve the space challenge is to break the list of obsolete locators into pieces so that a delayed transaction $T$ prevents from being reclaimed only the locator that $T$ has a direct reference as in the STMs using $CAS$. The idea is based on the fact that only the head of $O$'s locator-list is needed for further accesses to the $O$ object.

However, breaking the list of an obsolete object $O$ also creates another challenge on finding the head of $O$'s locator-list. Obviously, we cannot use a head pointer as in non-blocking linked-lists since modifying such a pointer requires $CAS$.
The key idea is to utilize the fact that there are no nested transactions and thus each thread has at most one {\em active} locator\footnote{An {\em active} locator is a locator that is still in use, opposite to an {\em obsolete} locator.} in each locator list. Therefore, by recording the latest locator of each thread appended to $O$'s locator-list, a transaction can find the head of $O$'s locator list. The solution is elaborated further in Section \ref{sec:Algorithm} and Section \ref{sec:Correctness}. 

Based on the key ideas, we come up with the data structure for a transactional memory object that is illustrated in Figure \ref{fig:NBFEB-STM} and presented in Algorithm \ref{alg:StartSTM}.  

The transactional memory object in NBFEB-STM is an array of $N$ pairs (pointer, timestamp), where $N$ is the number of concurrent threads/transactions as shown in Figure \ref{fig:NBFEB-STM}. Item $TMObj[i]$ is modified only by thread $t_i$ and can be read by all threads. Pointer $TMObj[i].loc$ points to the locator called $Loc_i$ corresponding to the latest transaction committed/aborted by thread $t_i$. Timestamp $TMObj[i].ts$ is the commit timestamp of the object referenced by $Loc_i.old$. 
After successfully appending its locator $Loc_i$ to  the list by executing $TFAS(head.next,Loc_i)$, $t_i$ will update its own item $TMObj[i]$ with its new locator $Loc_i$.
The $TMObj$ array is used to find the head of the list of locators $Loc_1, \cdots, Loc_N$.

For each locator $Loc_i$, in addition to fields $Tx, old$ and $new$ that reference the corresponding transaction object, the old data object and the new data object, respectively, as in DSTM\cite{HerLMS03}, there are two other fields $cts$ and $next$. The $cts$ field records the commit timestamp of the object referenced by $old$. The $next$ field is the pointer to the next locator in the locator list. The $next$ pointer is modified by NB-FEB primitives. In Figure \ref{fig:NBFEB-STM}, values $\{0,1\}$ in the $next$ pointer denote the values $\{false,true\}$ of its flag, respectively. The $next$ pointer of the head of the locator list, $Loc_3.next$, has its flag clear (i.e. 0), and the $next$ pointers of previous locators (e.g. $Loc_1.next$, $Loc_2.next$) have their flags set (i.e. 1) since their $next$ pointers were changed. The $next$ pointer of a new locator (e.g. $Loc_4.next$) is initialized to $(\perp,0)$.
Due to the garbage collector semantics, all locators $Loc_j$ reachable from the $TMObj$ {\em shared} object by following their $Loc_j.next$ pointers, will not be reclaimed.

For each transaction object $Tx_i$, in addition to fields $status$, $readSet$ and $writeSet$ corresponding to the status, the set of objects opened for read, and the set of objects opened for write, respectively, there is a field $cts$ recording $Tx_i$'s commit timestamp (if $Tx_i$ committed) as in LSA-STM \cite{RieFF06}. 

\subsection{Algorithm} \label{sec:Algorithm}

A thread $t_i$ starts a transaction $T$ by calling the {\sc StartSTM}$(T)$ procedure (Algorithm \ref{alg:StartSTM}). The procedure sets $T.status$ to $Active$ and clears its flag using $SAC$ (cf. Algorithm \ref{alg:SAC}). The procedure then initializes the lazy snapshot algorithm (LSA) \cite{RieFF06} by calling {\sc LSA\_Start}. NBFEB-STM utilizes LSA to preclude inconsistent views by {\em live} transactions, an essential aspect of transactional memory semantics \cite{GueK08}. The LSA has been shown to be an efficient mechanism to construct consistent snapshots for transactions \cite{RieFF06}. Moreover, the LSA can utilize up to $(N+1)$ versions of an transactional memory object $TMObj$ recorded in $N$ locators of $TMObj$'s locator list.
Note that the global counter $CT$ in LSA can be implemented by the {\em fetch-and-increment} primitive \cite{GotGKMRS82}, a combinable (and thus scalable) primitive \cite{KruRS88}. Except for the global counter $CT$, the LSA in NBFEB-STM does not need any strong synchronization primitives other than $TFAS$.  The {\sc Abort}$(T)$ operation in LSA, which is used to abort a transaction $T$, is replaced by $TFAS(T.status, Aborted)$. Note that the $status$ field is the only field of a transaction object $T$ that can be modified by other transactions.

\begin{algorithm}[t]
  \caption{{\sc StartSTM}($T$: transaction)} \label{alg:StartSTM}
	$TMObj$: array$[N]$ of $\{ptr, ts\}$. Pointer $TMObj[i].ptr$ points to the locator called $Loc_i$ corresponding to the latest transaction committed/aborted by thread $t_i$. Timestamp $TMObj[i].ts$ is the commit timestamp of the object referenced by $Loc_i.old$. $N$ is the number of concurrent threads/transactions. $TMObj[i]$ is written only by thread $t_i$. \\

	$Locator$: {\bf record} 
		$tx, new, old$: pointer;
		$cts$: timestamp; 
	{\bf end}. The $cts$ timestamp is the commit timestamp of the old version.\\

	$Transaction$: {\bf record}
		$status: \{Active, Committed, Aborted\}$;
		$cts$: timestamp; 
	{\bf end}. NBFEB-STM also keeps read/write sets as in LSA-STM, but the sets are omitted from the pseudocode since managing the sets in NBFEB-STM is similar to LSA-STM. \\


  	\algsetup{linenodelimiter=S:}
  	\begin{algorithmic}[1]
	\STATE {\sc SAC}$(T.status, Active)$; \COMMENT{Store-and-clear} \label{alg:S:initStatus}
	\STATE {\sc LSA\_Start}$(T)$ \COMMENT{Lazy snapshot algorithm}
  	\end{algorithmic}
\end{algorithm}

When a transaction $T$ opens an object $O$ for read, it invokes the {\sc OpenR} procedure (Algorithm \ref{alg:OpenRObject}). The procedure simply calls the {\sc LSA\_Open} procedure of LSA \cite{RieFF06} in the $Read$ mode to get the version of $O$ that maintains a consistent snapshot with the versions of other objects being accessed by $T$.
If no such a version of $O$ exists, {\sc LSA\_Open} will abort $T$ and consequently {\sc OpenR} will return $\perp$ (line \ref{alg:R:ReturnNil}R). That means there is a conflicting transaction that makes $T$ unable to maintain a consistent view of all the object being accessed by $T$. Otherwise, {\sc OpenR} returns the version of $O$ that is selected by LSA. This version is guaranteed by LSA to belong to a consistent view of all the objects being accessed by $T$.
Up to $(N+1)$ versions are available for each object $O$ in NBFEB-STM (cf. Lemma \ref{lem:numOfVersions}).
Since NBFEB-STM utilizes LSA, read-accesses to an object $O$ are invisible to other transactions and thus do not change $O$'s locator list.

\begin{algorithm}[t]
  \caption{{\sc OpenR}($T$: Transaction; $O_i$: TMObj): Open a transactional onject for read} \label{alg:OpenRObject}
  \algsetup{linenodelimiter=R:}
  \begin{algorithmic}[1]
	\ENSURE	reference to a {\em data} object if succeeds, or $\perp$.
	\STATE {\sc LSA\_Open}$(T, 0_i, "Read")$; \COMMENT{LSA's {\sc Open} procedure} \label{alg:R:LSAOpen}
	\IF{$T.status = Aborted$} \label{alg:R:TStatus}
	\RETURN $\perp$; \label{alg:R:ReturnNil}
	\ELSE
	\RETURN the version chosen by {\sc LSA\_Open}; \label{alg:R:ReturnVersion}
	\ENDIF
  	\end{algorithmic}
\end{algorithm}


When a transaction $T$ opens an object $O$ for write, it invokes the {\sc OpenW} procedure (cf. Algorithm \ref{alg:OpenWObject}). The task of the procedure is to append to the head of $O$'s locator list a new locator $L$ whose $Tx$ and $old$ fields reference to $T$ and $O$'s latest version, respectively. In order to find $O$'s latest version, 
the procedure invokes {\sc FindHead} (cf. Algorithm \ref{alg:FindHead}) to find the current head of $O$'s locator list (line \ref{alg:W:findHead}W). When the head called $H$ is found, the procedure determines $O$'s latest version based on the status of the corresponding transaction $H.Tx$ as in DSTM \cite{HerLMS03}. 
If the $H.Tx$ transaction committed, $O$'s latest version is $H.new$ with commit timestamp $H.Tx.cts$ (lines \ref{alg:W:ifCommit}W-\ref{alg:W:TxCts}W). A copy of $O$'s latest version is created and referenced by $L.new$ (line \ref{alg:W:CopyHNew}W) (cf. locators $Loc_2$ and $Loc_3$ in Figure \ref{fig:NBFEB-STM} as $H$ and $L$, respectively, for an illustration).
If the $H.Tx$ transaction aborted, $O$'s latest version is $H.old$ with commit timestamp $H.cts$ (lines \ref{alg:W:ifAbort}W-\ref{alg:W:HCts}W) (cf. locators $Loc_1$ and $Loc_2$ in Figure \ref{fig:NBFEB-STM} as $H$ and $L$, respectively, for an illustration). 
If the $H.Tx$ transaction is active, {\sc OpenW} consults the contention manager \cite{GeuHP05,SchS05} (line \ref{alg:W:CM}W) to solve the conflict between the $T$ and $H.Tx$ transactions. If $T$ must abort, {\sc OpenW} tries to change $T.status$ to $Aborted$ using $TFAS$ (line \ref{alg:W:AbortT}W) and returns $\perp$. Note that other transactions change $T.status$ only to $Aborted$, and thus if $TFAS$ at line \ref{alg:W:AbortT}W fails, $T.status$ has been changed to $Aborted$ by another transaction.
If $H.Tx$ must abort, {\sc OpenW} changes $H.Tx.status$ to $Aborted$ using $TFAS$ (line \ref{alg:W:AbortHTx}W) and checks $H.Tx.status$ again. 

The latest version of $O$ is then checked to ensure that it, together with the versions of other objects being accessed by $T$, belongs to a consistent view using {\sc LSA\_Open} with "Write" mode (line \ref{alg:W:LSAOpen}W). If it does, {\sc OpenW} tries to append the new locator $L$ to $O$'s locator list by changing the $H.next$ pointer to $L$ (line \ref{alg:W:TFASHead}W). Note that the $H.next$ pointer was initialized to $\perp$ with a clear flag, before $H$ was successfully appended to $O$'s locator list (line \ref{alg:W:initNext}W).
If {\sc OpenW} does not succeed, another locator has been appended as a new head and thus {\sc OpenW} must retry to find the new head (line \ref{alg:W:FindHeadAgain}W). Otherwise, it successfully appends the new locator $L$ as the new head of $O$'s locator list. {\sc OpenW}, which is being executed by a thread $t_i$, then makes $O[i].ptr$ reference to $L$ and records $L.cts$ in $O[i].ts$ (line \ref{alg:W:UpdateOi}W). This removes $O$'s reference to the previous locator $oldLoc$ appended by $t_i$, allowing $oldLoc$ to be reclaimed by the garbage collector. Since $oldLoc$ now becomes an obsolete locator, its $next$ pointer is reset (line \ref{alg:W:oldLocNext}W) to break possible chains of obsolete locators reachable by a delayed/halted thread, helping $oldLoc$'s descendant locators in the chains be reclaimed. 
For each item $j$ in the $O$ array such that $O[j].ts < O[i].ts$, the $O[j].ptr$ locator now becomes obsolete in a sense that it no longer keeps $O$'s latest version although it is still referenced by $O[j]$ (since only thread $t_j$ can modify $O[j]$). In order to break the chains of obsolete locators, {\sc OpenW} resets the $next$ pointer of the $O[j].ptr$ locator so that $O[j].ptr$'s descendant locators can be reclaimed by the garbage collector (lines \ref{alg:W:ClearOjLoop}W-\ref{alg:W:ClearOj}W). This chain-breaking mechanism makes the space complexity of an object updated by $N$ concurrent transactions/threads in NBFEB-STM be $\Theta(N)$, the optimal (cf. Theorem \ref{the:spaceComplexity}).

\begin{algorithm}[p!]
  \caption{{\sc OpenW}($T$: Transaction; $O$: TMObj): Open a transactional memory object for write by a thread $p_i$} \label{alg:OpenWObject}
  \algsetup{linenodelimiter=W:}
  \begin{algorithmic}[1]
	\ENSURE	reference to a {\em data} object if succeeds, or $\perp$.
	\STATE $newLoc \leftarrow$ new Locator; \label{alg:W:newLoc}
	\WHILE{\TRUE}	 	
	\STATE $head \leftarrow$ {\sc FindHead}$(O)$;  \COMMENT{Find the head of $O$'s list.} \label{alg:W:findHead}
	\FOR{$i=0$ to $1$}	
 		\IF{ $head.tx.status = Committed$} \label{alg:W:ifCommit}
			\STATE $newLoc.old \leftarrow head.new$; \label{alg:W:HNew} 
			\STATE $newLoc.cts \leftarrow head.tx.cts$; \label{alg:W:TxCts}
			\STATE $newLoc.new \leftarrow$ {\sc Copy}$(head.new)$;\COMMENT{Create a duplicate} \label{alg:W:CopyHNew}
			\STATE {\bf break};
		\ELSIF{ $head.tx.status = Aborted$} \label{alg:W:ifAbort}
			\STATE $newLoc.old \leftarrow head.old$; \label{alg:W:HOld}
			\STATE $newLoc.cts \leftarrow head.cts$; \label{alg:W:HCts}
			\STATE $newLoc.new \leftarrow$ {\sc Copy}$(head.old)$; \label{alg:W:CopyHOld}
			\STATE {\bf break};
		\ELSE 
			\STATE $myProgession \leftarrow$ {\sc CM}$(O_i, "Write")$\COMMENT{ $head.tx$ is active $\Rightarrow$ Consult the contention manager} \label{alg:W:CM}
			\IF{ $myProgression =$ \FALSE}
				\STATE {\sc TFAS}$(T.status, Aborted)$; \COMMENT{If fails, another has executed this $TFAS$.} \label{alg:W:AbortT}
				\RETURN $\perp$;
			\ELSE  
				\STATE {\sc TFAS}$(head.tx.status, Aborted)$; \label{alg:W:AbortHTx}  
				\STATE {\bf continue}; \COMMENT{ Transaction $head.tx$ has committed/aborted $\Rightarrow$ Check $head.tx.status$ one more time}
			\ENDIF
		\ENDIF \label{alg:W:endIf}
	\ENDFOR
	\STATE $newLoc.tx \leftarrow T$; 
	\STATE {\sc SAC}$(newLoc.next, \perp)$; \COMMENT{Store-and-clear} \label{alg:W:initNext}
	\STATE {\sc LSA\_Open}$(T, O, "Write")$; \COMMENT{LSA's {\sc Open} procedure.} \label{alg:W:LSAOpen}
	\IF{$T.status = Aborted$}
	\RETURN $\perp$; \COMMENT{Performance (not correctness): Don't add $newLoc$ to $O$ if $T$ has aborted due to, for instance, {\sc LSA\_Open}.}
	\ENDIF
	\IF{ {\sc TFAS}$(head.next, newLoc) \neq \perp$} \label{alg:W:TFASHead}
		\STATE {\bf continue}; \COMMENT{Another locator has been appended $\Rightarrow$ Find the head again} \label{alg:W:FindHeadAgain}
	\ELSE
		\STATE $oldLoc = O[i]$;
		\STATE $O[i] \leftarrow (newLoc, newLoc.cts)$; \COMMENT{Atomic assignment; $p_i$'s old locator is unlinked from $O$.} \label{alg:W:UpdateOi} 
		\STATE {\sc SAC}$(oldLoc.next, \perp)$; \COMMENT{$oldLoc$ may be in the chain of a sleeping thread $\Rightarrow$ Stop the chain here} \label{alg:W:oldLocNext}
		\FOR{ each item $L_j$ in $O$ such that $L_j.ts < O[i].ts$} \label{alg:W:ClearOjLoop}
			\STATE {\sc SAC}$(L_j.ptr.next, \perp)$ \COMMENT{Reset the $next$ pointer of the obsolete locator} \label{alg:W:ClearOj}
		\ENDFOR
		\RETURN $newLoc.new$;
	\ENDIF
	\ENDWHILE

  	\end{algorithmic}
\end{algorithm}

In order to find the head of $O$'s locator list as in {\sc OpenW}, a transaction invokes the {\sc FindHead}$(O)$ procedure (cf. Algorithm \ref{alg:FindHead}). The procedure atomically reads $O$ into a local array $start$ (line \ref{alg:F:scanO}F). Such a multi-word read operation is supported by emerging multicore architectures like CUDA \cite{CUDA} and Cell BE \cite{GscHFHWY06}. In the contemporary chips of these architectures, a read operation can atomically read 128 bytes. In general, such a multi-word read operation can be implemented as an atomic snapshot using only single-word read and single-word write primitives \cite{AfeADGMS93}. {\sc FindHead} finds the item $start_{latest}$ with the highest timestamp in $start$ and searches for the head from locator $start_{latest}.ptr$ by following the $next$ pointers until it finds a locator $H$ whose $next$ pointer is $\perp$ (lines \ref{alg:F:latest}F-\ref{alg:F:tmpNext}F). Since some locators may become obsolete and their $next$ pointers were reset to $\perp$ by concurrent transactions (lines \ref{alg:W:oldLocNext}W and \ref{alg:W:ClearOj}W in Algorithm \ref{alg:OpenWObject}), {\sc FindHead} needs to check $H$'s commit timestamp against the highest timestamp of $O$ at a moment after $H$ is found (lines \ref{alg:F:scanO2}F-\ref{alg:F:until}F). If $H$'s commit timestamp is greater than or equal to the highest timestamp of $O$, $H$ is the head of $O$'s locator list (cf. Lemma \ref{lem:FindHead}). Otherwise, $H$ is an obsolete locator and {\sc FindHead} must retry (line \ref{alg:F:until}F). The {\sc FindHead} procedure is lock-free, namely it will certainly return the head of $O$'s  locator list after at most $N$ iterations unless a concurrent thread has completed a transaction and subsequently has started a new one, where $N$ is the number of concurrent (updating) threads (cf. Lemma \ref{lem:FindHead_LF}).
Note that as soon as a thread obtains $head$ from {\sc FindHead} (line \ref{alg:W:findHead}W of {\sc OpenW}, Algorithm \ref{alg:OpenWObject}), the locator referenced by $head$ will not be reclaimed by the garbage collector until the thread returns from the {\sc OpenW} procedure.

\begin{algorithm}[t]
  \caption{{\sc FindHead}($O$: TMObj): Find the head of the locator list} \label{alg:FindHead}
  \algsetup{linenodelimiter=F:}
  \begin{algorithmic}[1]
	\ENSURE reference to the head of the locator list
	\REPEAT
		\STATE $start \leftarrow O$; \COMMENT{Read $O$ to a local array atomically.} \label{alg:F:scanO}
		\STATE Let $start_{latest}$ is the item with highest timestamp; \label{alg:F:latest}
		\STATE $tmp \leftarrow start_{latest}.ptr$; \COMMENT{Find a locator whose $next$ pointer is $\perp$}
		\WHILE{ $tmp.next \neq \perp$} \label{alg:F:while}
			\STATE $tmp \leftarrow tmp.next$; \label{alg:F:tmpNext}
		\ENDWHILE
		\STATE $start' \leftarrow O$; \COMMENT{Check if $tmp$ is the head.} \label{alg:F:scanO2}
		\STATE Let $start'_{latest}$ is the item with highest timestamp;
	\UNTIL{ $tmp.cts \geq start'_{latest}.ts$}; \label{alg:F:until}
	\RETURN $tmp$;
  	\end{algorithmic}
\end{algorithm}

When committing, read-only transactions in NBFEB-STM do nothing and always succeed in their commit phase as in LSA-STM \cite{RieFF06}. They can abort only when trying to open an object for read (cf. Algorithm \ref{alg:OpenRObject}). Other transactions $T$, which have opened at least one object for write, invoke the {\sc CommitW} procedure (Algorithm \ref{alg:Commit}). The procedure calls the {\sc LSA\_Commit} procedure to ensure that $T$ still maintains a consistent view of objects being accessed by $T$ (line \ref{alg:C:LSACommit}C). $T$'s commit timestamp is updated with the timestamp returned from {\sc LSA\_Commit} (line \ref{alg:C:TCts}C). Finally, {\sc CommitW} tries to change $T.status$ to $Committed$ (line \ref{alg:C:TFAS}C). $T.status$ will be changed to $Committed$ at this step if it has not been changed to $Aborted$ due to the semantics of $TFAS$.

\begin{algorithm}[t]
  \caption{{\sc CommitW}($T$: Transaction): Try to commit an update transaction $T$ by thread $p_i$} \label{alg:Commit}
  \algsetup{linenodelimiter=C:}
  \begin{algorithmic}[1]
	\STATE $CT_{T} \leftarrow$ {\sc LSA\_Commit}$(T)$; \COMMENT{Check consistent snapshot. $CT_{T}$ is $T$'s unique commit timestamp from LSA.} \label{alg:C:LSACommit}
	\STATE $T.cts \leftarrow CT_{T}$; \COMMENT{Commit timestamp of $T$ if $T$ manages to commit.} \label{alg:C:TCts}
	\STATE {\sc TFAS}$(T.status, Committed)$; \label{alg:C:TFAS}
  	\end{algorithmic}
\end{algorithm}

\subsection{Analysis} \label{sec:Correctness}

In this section, we prove that NBFEB-STM fulfills the three essential aspects of 
transactional memory semantics \cite{GueK08}:
\begin{description}
\item[Instantaneous commit]: Committed transactions must appear as if they executed instantaneously at some unique point in time, and aborted transactions, as if they did not execute at all.
\item[Preserving real-time order]: If a transaction $T_i$ commits before a transaction $T_j$ starts, then $T_i$ must appear as if it executed before $T_j$. 
Particularly, if a transaction $T_1$ modifies an object $O$ and commits, and then another transaction $T_2$ starts and reads $O$, then $T_2$ must read the value written by $T_1$ and not an older value.
\item[Preluding inconsistent views]: The state (of shared objects) accessed by {\em live} transactions must be consistent.
\end{description}

First, we prove some key properties of NBFEB-STM.

\begin{lemma} \label{lem:nextPointer}
A locator $L_i$ with timestamp $cts_i$ does not have any links/references to another locator $L_j$ with a lower timestamp $cts_j < cts_i$. 
\end{lemma}
\begin{proof}
There is only the $next$ pointer to link between locators. 
The $next$ pointer of locator $L_i$ points to a locator $L_j$ only if $L_j.cts$ is not less than $L_i.cts$ (lines \ref{alg:W:TxCts}W and \ref{alg:W:HCts}W, Algorithm \ref{alg:OpenWObject}). Note that for each locator $L_i$, the commit timestamp $L_i.tx.cts$ of its corresponding transaction $L_i.tx$ (if $L_i.tx$ committed) is the commit timestamp of $L$'s new data and thus it is always greater than the commit timestamp $L_i.cts$ of $L_i$'s old data. 
\end{proof}

\begin{lemma}\label{lem:FindHead}
The locator returned by {\sc FindHead}$(O)$ (Algorithm \ref{alg:FindHead}) is the head $H$ of $O$'s locator list at the time-point {\sc FindHead} found $H.next = \perp$ (line \ref{alg:F:while}F).
\end{lemma}
\begin{proof}
Let $L$ be the locator returned by {\sc FindHead}. 
Since the $next$ pointer of a new locator is initialized to $\perp$ (line \ref{alg:W:initNext}W, Algorithm \ref{alg:OpenWObject}) before the locator is appended into the list by $TFAS$  (line \ref{alg:W:TFASHead}W), {\sc FindHead} will find a locator $L$ whose $next$ pointer is $\perp$ at a time-point $tp$ (line \ref{alg:F:while}F). The $L$ locator is either the head at that time or a reset locator (due to lines \ref{alg:W:oldLocNext}W and \ref{alg:W:ClearOj}W, Algorithm \ref{alg:OpenWObject}).

If $L$ is a reset locator, $start'_{latest}.cts > L.cts$ holds (line \ref{alg:F:until}F) since a locator is reset (e.g. $oldLoc$ at line  \ref{alg:W:oldLocNext}W or $L_j$ at line \ref{alg:W:ClearOj}W) only after a locator with a higher timestamp (e.g. $newLoc$) has been written into the $O$ array (line \ref{alg:W:UpdateOi}W). Since {\sc FindHead} atomically reads the $O$ array after it found $L.next = \perp$, it will observe the higher timestamp.
This makes {\sc FindHead} retry and discard $L$, a contradiction to the hypothesis that $L$ is returned by {\sc FindHead}. Therefore, the $L$ locator returned by {\sc FindHead} must be the head at the time-point {\sc FindHead} found $L.next = \perp$ (line \ref{alg:F:while}F).
\end{proof}

Since a thread must get a result from {\sc FindHead} (line \ref{alg:W:findHead}W) before it can consult the contention manager (line \ref{alg:W:CM}W), {\sc FindHead} must be lock-free (instead of being obstruction-free) in order to guarantee the obstruction-freedom for transactions.

\begin{lemma}\label{lem:FindHead_LF}
{\em (Lock-freedom)} {\sc FindHead}$(O)$ will certainly return the head of $O$'s  locator list after at most $N$ repeat-until iterations unless a concurrent thread has completed a transaction and subsequently has started a new one, where $N$ is the number of concurrent threads updating $O$. 
\end{lemma}
\begin{proof}
From Lemma \ref{lem:FindHead}, any locator returned by {\sc FindHead}$(O)$ is the head of $O$'s locator list. Therefore, we only need to prove that {\sc FindHead}$(O)$ will certainly return a locator after at most $N$ iterations unless a concurrent thread has completed a transaction and subsequently has started a new one.

We prove this by contradiction. Assume that {\sc FindHead}$(O)$ executed by thread $t_i$, does not return after $N$ iterations and no thread has completed its transaction since {\sc FindHead} started.
Since each thread $t_j$ updates its own item $O[j]$ only once when opening $O$ for update (line \ref{alg:W:UpdateOi}W, , Algorithm \ref{alg:OpenWObject}), at most $(N-1)$ items $j$ of $O, j \neq i,$ have been updated since {\sc FindHead}$(O)$ started.

First we prove that {\sc FindHead}$(O)$ will return in the iteration during which no item of $O$ is updated between the first atomic read (line \ref{alg:F:scanO}F) and the second atomic read of the $O$ array (line \ref{alg:F:scanO2}F). 

Indeed, since each transaction successfully appends its own locator to the head of $O$'s locator list only once when opening $O$ for update (line \ref{alg:W:TFASHead}W), at most $(N-1)$ locators are appended to $O$'s locator list after the first scan. Therefore, {\sc FindHead} will certainly find a locator $L$ such that $L.next \neq \perp$ (line \ref{alg:F:while}F) in the current repeat-until iteration. Note that for each $next$ pointer, only the first transaction executing $TFAS$ on the pointer, manages to append its locator to the pointer.

Since (1) the $next$ pointer of a locator $L_i$ points to a locator $L_j$ only if $L_j.cts \geq L_i.cts$ (cf. Lemma \ref{lem:nextPointer}) and (2) {\sc FindHead} found $L$ by following the $next$ pointers starting from $start_{latest}.ptr$ (lines \ref{alg:F:latest}F-\ref{alg:F:tmpNext}F), we have $L.cts \geq start_{latest}.ptr.cts$. Note that $start_{latest}.ptr.cts = start_{latest}.ts$ (line \ref{alg:W:UpdateOi}W). 
Since no item of $O$ is updated between the first scan (line \ref{alg:F:scanO}F) and the second scan of the $O$ array (line \ref{alg:F:scanO2}F), the items with highest timestamp of both scans are the same, i.e. $start_{latest} = start'_{latest}$. Therefore, $L.cts \geq start'_{latest}.ts$ holds (line \ref{alg:F:until}F) and $L$ is returned. 

Since {\sc FindHead} executed by thread $t_i$ does not return after $N$ iterations due to hypothesis, it follows that at least $N$ items have been updated since {\sc FindHead} started, a contradiction to the above argument that at most $(N-1)$ items have been updated since {\sc FindHead} started.
\end{proof}

\remove {
Question: 
When to update O[i] (by $p_i$)? Successfully insert a new locator or Commit/Abort time? 

- If update occurs at Abort time, O[i] must be writable by other threads since other threads allow to abort $p_i$. 

- If update occurs only at Commit time (but not Abort time), the number of locators of the list would be at least $O(N^2)$ since one committed Tx can abort all other Tx, forcing them to retry. The number of aborted locators in the list is (N-1) + (N-2) + (N-3) + ... + 1 = O(N^2) 

- Therefore, update occurs at a successful insertion of new locator is the best since it keeps the number of locator in the list is at most $2N$: $p_i$'s old locator is removed from the list as soon as the new locator is added.
} 


\begin{lemma}
{\em (Instantaneous commit)} TFAS-LSA guarantees that committed transactions appear as if they executed instantaneously and aborted transactions appear as if they did not execute at all.  
\end{lemma}
\begin{proof}
Similar to the DSTM \cite{HerLMS03} and LSA-STM \cite{RieFF06}, the NBFEB-STM uses the indirection technique that allows a transaction $T_j$ to  commit its modifications to all objects in its write-set instantaneously by switching its status from $Active$ to $Committed$. Its committed status must no longer be changed. NBFEB-STM uses the $TFAS$ primitive (Algorithm \ref{alg:TFAS}) to achieve the property (line \ref{alg:C:TFAS}C, Algorithm \ref{alg:Commit}). 
Since the flag of the $T_j.status$ variable is $false$ (or 0) when the transaction starts (line \ref{alg:S:initStatus}S, Algorithm \ref{alg:StartSTM}), only the first $TFAS$ primitive can change the variable.
If $T_j$ manages to change the $T_j.status$ variable to $Committed$, the  variable is no longer able to be changed using $TFAS$ until the transaction object $T_j$ is reclaimed by the garbage collector. Note that even if thread $t_j$ completed transaction $T_j$ and has started another transaction $T'_j$, the transaction object $T_j$ will not be reclaimed until all the locators keeping a reference to $T_j$ are reclaimable.

Since active transactions $T_j$ make all changes on their own copy $T_j.new$ of a shared object $O$ before their status is changed from $Active$ to either $Aborted$ or $Committed$, aborted transactions do not affect the value of $O$.
\end{proof}

The two other correctness criteria for transactional memory are precluding inconsistent views and preserving real-time order \cite{GueK08}. 
Since TFAS use the lazy snapshot algorithm $LSA$ \cite{RieFF06}, the former will follow if we can prove that the LSA algorithm is integrated correctly into NBFEB-STM.

\begin{lemma}\label{lem:LSACorrectness}
The versions kept in $N$ locators $O[j].ptr, 1 \leq j \leq N$, for each object $O$ is enough for checking the validity of a transaction $T$ using the LSA algorithm \cite{RieFF06}, from the correctness point of view.
\end{lemma}

\begin{proof}
The LSA algorithm requires only the commit timestamp (i.e. $\lfloor O^{CT} \rfloor$ \footnote{Term $\lfloor O^{t} \rfloor$ denotes the time of most recent update of object $O$ performed no later than time $t$ \cite{RieFF06}.}) of the most recent version (i.e. $O^{CT}$ \footnote{Term $O^{t}$ denotes the content/version of object $O$ at time $t$ \cite{RieFF06}.}) of each object $O$ at a timestamp $CT$ when it checks the validity of a transaction $T$.
The older versions of $O$ are not required for correctness - they only increase the chance that a suitable object version is available.

We will prove that by atomically reading the $O$ object/array at the timestamp $CT$ to a local variable $V$  as at line \ref{alg:F:scanO}F in Algorithm \ref{alg:FindHead}, LSA will find the commit timestamp $\lfloor O^{CT} \rfloor$. 


A new version of $O$ is created and becomes accessible by all transactions when a transaction $T_j$ commits its modification $L_j.new$ (stored in locator $L_j$) to $O$ by changing its status from $Active$ to $Committed$ (line \ref{alg:C:TFAS}C, Algorithm \ref{alg:Commit}). Since every transaction $T_j$ writes its locator $L_j$ to $O[j].ptr$ when opening $O$ for update (line \ref{alg:W:UpdateOi}W, Algorithm \ref{alg:OpenWObject}) (i.e. before committing), at least one of the locators $O[j].ptr, 1 \leq j \leq N$, must contain the most recent version of $O$ at the timestamp $CT$ when $O$ is read to $V$.

Since a transaction $T_j$ updates $O[j]$ with its new locator $L_j$ only after 
 successfully appending $L_j$ to the head of $O$'s locator list, at most one of the locators $O[j].ptr, 1 \leq j \leq N,$ is the head of the list at the timestamp $CT$ when the snapshot $V$ of $O$ is taken. Other locators $V[j].ptr$ that are not the head, have their transactions committed/aborted before $CT$. Note that as soon as the transaction of a locator committed/aborted, the locator's versions together with their commit timestamp is no longer changed.
If transaction $V[i].ptr.tx$ committed, the version kept in locator $V[j].ptr$ is $V[j].ptr.new$ with commit timestamp $V[j].ptr.tx.cts$, the commit timestamp of the transaction. If transaction $V[j].ptr.tx$ has been aborted or is active, the version is $V[j].ptr.old$ with commit timestamp $V[j].ptr.cts$. The only 
possible version with commit timestamp higher than $CT$ is $V[h].ptr.new$ where $V[h].ptr$ was the head at the timestamp $CT$ when $V$ was taken and then transaction $V[h].ptr.tx$ committed. In this case, $V[h].ptr.old$ is the most recent version at $CT$ and its commit timestamp is $V[h].ptr.cts$.

Therefore, by checking the commit timestamps of the versions kept in each locator $V[j].ptr, 1 \leq j \leq N,$ against $CT$, LSA will find the commit timestamp $\lfloor O^{CT} \rfloor$ of the most recent update of object $O$ performed no later than $CT$.





\end{proof}

\begin{lemma} \label{lem:numOfVersions}
The number of versions available for each object in NBFEB-STM is up to $(N+1)$, where $N$ is the number of threads.
\end{lemma}
\begin{proof}
For each object $O$, each thread $t_j$ keeps a version of $O$ that has been  accessed most recently by $t_j$, in locator $O[j].ptr$ (or $L_j$ for short). If $t_j$'s latest transaction $T_j$ committed $\forall j \in [1,N]$, the $L_j.old$ is an old version of $O$ with validity range $[L_j.cts, L_j.tx.cts)$ 
\footnote{The {\em validity range} of a version $v_i$ of an object $O$ is the interval from the commit time of $v_i$ to the commit time of the next version $v_{i+1}$ of $O$ \cite{RieFF06}.}. 
Therefore, if every thread has its latest transaction committed, each object $O$ updated by $N$ threads will have $N$ old versions with validity ranges, additional to its latest version.  
\end{proof}

\begin{lemma}
{\em (Consistent view)} NBFEB-STM precludes inconsistent views of shared objects from {\em live} transactions.
\end{lemma}
\begin{proof}
Since the LSA lazy snapshot algorithm is correctly integrated into NBFEB-STM (Lemma \ref{lem:LSACorrectness}), the lemma follows.   
\end{proof}

\begin{definition}
The {\em value} of a locator $L$ is either $L.new$ if $L.tx.status=Committed$, or $L.old$ otherwise.
\end{definition}

\begin{lemma}\label{lem:LocValue}
In each $O$'s locator list, the old value $L'.old$ of a locator $L'$ is not older than the value of its previous locator \footnote{A locator $L$ is a {\em previous} locator of a locator $L'$ if starting from $L$ we can reach $L'$ by following $next$ pointers.} $L$. 
\end{lemma}
\begin{proof}
Let $L''$ be the locator pointed by $L.next$. Since $L.tx.status$ must be either $Committed$ or $Aborted$ (but not $Active$) before $L''$ is appended to $L.next$ (lines \ref{alg:W:ifCommit}W-\ref{alg:W:endIf}W, Algorithm \ref{alg:OpenWObject}), $L''.old$ is $L$'s value, which is either $L.new$ if $L.tx.status=Committed$ (line \ref{alg:W:HNew}W) or $L.old$ if $L.tx.status=Aborted$ (line \ref{alg:W:HOld}W). That means $L''.old$ is not older than $L$'s value. Arguing inductively for all locators on the directed path from $L$ to $L'$, the lemma follows.
\end{proof}

\begin{lemma}
{\em (Real-time order preservation)} NBFEB-STM preserves the real-time order of transactions.
\end{lemma}
\begin{proof}

We need to prove that if a transaction $T_1$ modifies an object $O$ and commits and then another transaction $T_2$ starts and reads $O$, $T_2$ must read the value written by $T_1$ and not an older value \cite{GueK08}. Namely, $T_1$ is the most recent transaction committing its modification to $O$ before $T_2$ reads $O$.


First we prove that $T_2$ reads the value $v_1$ written by $T_1$ if $T_2$ opens $O$ for read (cf. {\sc OpenR}, Algorithm \ref{alg:OpenRObject}).
In the proof of Lemma \ref{lem:LSACorrectness}, we have proven that the value of $O$ read at a timestamp $CT$ by LSA is the most recent value of $O$ at that timestamp. Since $T_1$ is the most recent transaction committing its modification to $O$ before $T_2$ reads $O$, $v_1$ is in the set of available versions of $O$ read by {\sc LSA\_Open} (line \ref{alg:R:LSAOpen}R).  Since $T_1$ commits before $T_2$ starts and reads $O$, the commit timestamp of $v_1$ is less than the upper bound of any validity range $R_{T_2}$\footnote{The {\em validity range} $R_T$ of a transaction $T$ is the time range during which each of the objects accessed by $T$ is valid \cite{RieFF06}.}
 chosen by the {\sc LSA\_Open} (i.e. $\lfloor O^{CT} \rfloor \leq T_{max}$ in terminology used by LSA \cite{RieFF06}.) Therefore, the {\sc LSA\_Open} in {\sc OpenR} will return $v_1$, which is subsequently returned by {\sc OpenR} (line \ref{alg:R:ReturnVersion}R)

We now prove that $T_2$ reads the value $v_1$ written by $T_1$ if $T_2$ opens $O$ for read (cf. {\sc OpenW}, Algorithm \ref{alg:OpenWObject}). Particularly, we prove that the $old$ value of $T$'s new locator (lines \ref{alg:W:HNew}W and \ref{alg:W:HOld}W) is $v_1$.

Let $p_1$ and $p_2$ be the threads executing $T_1$ and $T_2$, respectively, $L_1$ be the locator containing $T_1$'s modification (in $L_1.new$) that is committed to $O$ and $v_2$ be the value of $O$ read by $T_2$.   
The $v_2$ value is the value of the head $H$ of $O$'s locator list returned from {\sc FindHead} executed by $T_2$, which is either $H.new$ if $H.ts.status=Committed$ or $H.old$ otherwise (line \ref{alg:W:HNew}W or \ref{alg:W:HOld}W). 

Since $T_1$ committed before $T_2$ started, $H$ is the head of $O$'s locator list that includes $L_1$ (cf. Lemma \ref{lem:FindHead}). Note that since $T_1$ is the latest transaction committing its modification to $O$,  all locators $L'$ that have ever been reachable from $L_1$ via $next$ pointers, have the most recent timestamp/value (cf. Lemma \ref{lem:LocValue}) and thus will not be reset (lines \ref{alg:W:ClearOjLoop}W-\ref{alg:W:ClearOj}W, Algorithm \ref{alg:OpenWObject}). Since there is a directed path from $L_1$ to $H$ via $next$ pointers, it follows from Lemma \ref{lem:LocValue} that the value of $H$ is not older than that of $L_1$. 

On other hand, since $T_1$ is the latest transaction committing its modification to $O$ before $T_2$ reads $O$, there is no value of $O$ that is newer than that of $L_1$. Therefore, the value of $H$ is the value of $L_1$. That means $T_2$ reads the $v_1$ value written by $T_1$.

Finally, we need to prove that {\sc LSA\_Open} at line \ref{alg:W:LSAOpen}W accepts $v_1$. Indeed, since $v_1$ is the most recent update of $O$ and $T_1$ commits before $T_2$ starts, the commit timestamp of $v_1$ is less than the upper bound of any validity range $R_{T_2}$ chosen by the {\sc LSA\_Open} (i.e. $\lfloor O^{CT} \rfloor \leq T_{max}$). Therefore, the {\sc LSA\_Open} at line \ref{alg:W:LSAOpen}W accepts $v_1$.
\end{proof}

\begin{lemma}\label{lem:numOfLocators}
For each object $O$, there are at most $4N$ locators that cannot be reclaimed by the garbage collector at any time-point, where $N$ is the number of update threads.
\end{lemma}
\begin{proof}
Let $L_i$ be a locator created by a thread $p_i$.
A locator $L_i$ cannot be reclaimed by the garbage collector if it is reachable by a thread. In NBFEB-STM, a locator $L_i$ is reachable if it is i) $p_i$'s {\em new} locator $newLoc$, ii) $p_i$'s {\em shared} locator, which is referenced directly by $O[i].ptr$, and iii) $p_i$'s {\em old} locators $oldLoc$ that is reachable by other threads. $p_i$'s shared locator will become one of $p_i$'s old locators if $O[i].ptr$ is updated with $p_i$'s new locator (line \ref{alg:W:UpdateOi}W, Algorithm \ref{alg:OpenWObject}). At that moment, $p_i$'s new locator becomes $p_i$'s shared locator. If there is no thread keeping a direct/indirect reference to $p_i$'s old locators, these locators are ready to be reclaimed (i.e. unreachable) when $p_i$ returns from the {\sc OpenW} procedure.

Let $C^p_i$ and $C^o_i$ be the chains of locators (linked by their $next$ pointers) that cannot be reclaimed due to thread $p_i$ and $O[i]$, respectively. The $C^p_i$ chain starts at the locator that is referenced directly by $p_i$ (not directly by $O$) and ends at either the locator whose $next$ pointer is $\perp$ or the locator whose next locator is referenced directly by another thread or $O$. The $C^o_i$ chain starts at the locator that is referenced directly by $O[i]$ and ends at either the locator whose $next$ pointer is $\perp$ or the locator that is referenced directly by another thread or $O$.
Note that there are no two locators whose $next$ pointers point to the same locator $L_j$ since $p_j$ successfully appends $L_j$ into the head of the locator list only once (line \ref{alg:W:TFASHead}W, Algorithm \ref{alg:OpenWObject}). 

At any time, each thread $p_i$ has at most one $C^p_i$ and one $C^o_i$. The $C^p_i$ starts either with $p_i$'s new locator (before assignment $O[i] \leftarrow newLoc$ at line \ref{alg:W:UpdateOi}W, Algorithm \ref{alg:OpenWObject}) or with $p_i$'s old locator (after this assignment). Since $p_i$ has a unique item in the $O$ array, it has at most one $C^o_i$. Therefore, there are at most $2N$ chains.

We will prove that if $p_i$ has three locators participating in chains (of arbitrary threads), at least one of the three locators must be the end-locator of a chain. Indeed, during the execution of the {\sc OpenW} procedure (Algorithm \ref{alg:OpenWObject}), $p_i$ creates only one new locator (line \ref{alg:W:newLoc}W) in addition to its locator $O[i].ptr$, if any. If $p_i$ has three locators that are participating in chains, at least one of them is $p_i$'s old locator $L^o$ resulting from one of $p_i$'s previous executions $E$ of {\sc OpenW}. Since $p_i$ sets the $next$ pointer of its old locator $oldLoc$ to $\perp$ before returning from $E$ (line \ref{alg:W:oldLocNext}W), $L^o$'s $next$ pointer is $\perp$. That means $L^o$ is the end-locator of a chain. 

It then follows that each thread has at most two {\em non-end} locators participating in all the chains. The number of non-end locators in all the chains is at most $2N$. Since there are at most $2N$ chains, there are at most $2N$ {\em end}-locators. Therefore, the total number of locators in all the chains is $4N$.
\end{proof}

\begin{theorem} \label{the:spaceComplexity}
{\em (Space complexity)} The space complexity of an object updated by $N$ threads in NBFEB-STM is $\Theta(N)$, the optimal.
\end{theorem}
\begin{proof}
Since each object $O$ in NBFEB-STM is an array of $N$ items (cf. Algorithm \ref{alg:StartSTM}), the space complexity of an object is $\Omega(N)$.

From Lemma \ref{lem:numOfLocators}, for each object $O$ there are at most $4N$ locators that cannot be reclaimed by the garbage collector at any point in time. Since each locator $L$ references to at most one transaction object $L.tx$ (cf. Figure \ref{fig:NBFEB-STM}), the space complexity of an object is $O(N)$. 

Due to the {\em instantaneous commit} requirement of transactional memory semantics \cite{GueK08}, when opening an object for update, each thread/transaction in any STM system must create a copy of the original object. Therefore, the space complexity of an object updated by $N$ threads is $O(N)$ for all STM systems. It follows that the space complexity $\Theta(N)$ of  an object updated by $N$ threads in NBFEB-STM is optimal.
\end{proof}

\begin{definition}
Contention level $CL_{l,t}$ of a memory location $l$ at a timestamp $t$ is the number of requests that need to be executed sequentially on the location by a memory controller (i.e. the number of requests for $l$ buffered at time $t$).
\end{definition}

\begin{definition}
Contention level of a transaction $T$ that starts at timestamp $s_T$ and ends (i.e. commits or aborts) at timestamp $e_T$ is $max_{s_T \leq t \leq e_T} CL_{l,t}$ for all memory locations $l$ accessed by $T$
\end{definition}

\begin{lemma}
{\em (Contention reduction)} Transactions using NBFEB-STM have lower contention levels than those using $CAS$-based STMs do.
\end{lemma}
\begin{proof}





({\it Sketch}) 
Since $CAS$ is not {\em combinable} \cite{KruRS88,BleGV08}, $M$ conflicting $CAS$ primitives on the same synchronization variable, like $TMObj$ pointer or a transaction's $status$ variable in $CAS$-based STMs \cite{HerLMS03, MarSS05, RieFF06}, issue $M$ remote-memory requests to the corresponding memory controller. Since $TFAS$ is combinable, the remote-memory requests from $M$ conflicting $TFAS$ primitives to the same variable, like the $next$ pointer or a transaction's $status$ variable in NBFEB-STM, can be combined into only one request to the corresponding memory controller. Therefore, the combinable primitive significantly reduces the number of requests for each memory location buffered at the memory controller.



\end{proof}

\section{Garbage Collectors} \label{sec:GarbageCollector}

In this section, we present a non-blocking garbage collection algorithm called NB-GC that can be used in the context of NBFEB-STM. The NB-GC algorithm does not requires synchronization primitives other than reads and writes while it still guarantees the obstruction-freedom property for {\em application threads} (or mutators in the memory management terminology). The obstruction-freedom here means that a halted application-thread cannot prevent other application-threads from making progress. 
 
Like previous concurrent garbage collection algorithms for multiprocessors \cite{Appel04, AzaLPP03, BacALRS01, BoeDS91, CheB01, DijLMSS78, DolL93, DolG94, DomKP00, Lamport76, LevP06, SinBWC07, SomDK06, SomK07, Steele75}, the new NB-GC algorithm is a priority-based garbage collection algorithm in which the collector thread is a privileged thread that may suspend and subsequently resume the mutator threads. 
The NB-GC algorithm is an improvement of the seminal on-the-fly garbage collector \cite{DijLMSS78, DolG94, DolL93} using the sliding view technique \cite{LevP06} called SV-GC. Unlike the SV-GC algorithm, the NB-GC algorithm allows the collector to suspend a mutator at any point in the mutator's code (even in the reference slot update and object allocation procedures). This prevents a mutator from blocking the collector and consequently from blocking other mutators.



In the concurrent garbage collection model, there are two kind of threads: application threads (e.g. the mutators) that perform user programs (error-prone codes), and privileged threads with higher priority (e.g. the collector) that perform system tasks (error-free codes). Whereas the application threads can be delayed/preempted arbitrarily, the system threads when running will not be preempted by the application threads. NB-GC guarantees obstruction-freedom for {\em application threads}, which usually perform users error-prone codes. Namely, a halted application-thread will not prevent other application-threads from making progress via blocking the garbage collector. 
The model, in some sense, covers the non-blocking garbage collection algorithms  \cite{HerLMM05, Mic04}  that, at the first look, seem not to require privileged threads. In fact, the non-blocking garbage collectors require strong synchronization primitives like {\em compare-and-swap} whose atomicity is guaranteed by hardware threads, a kind of privileged threads.

The SV-GC algorithm using the sliding view technique \cite{LevP06} does not need synchronization primitives other than reads and writes. However, it requires that the mutator be suspended only at a safe point, particularly it requires that the mutator not be stopped during the execution of a reference slot update nor new object allocation. If a mutator $M$ is preempted during such an execution, the collector cannot progress since it cannot suspend the mutator $M$. This would prevent the other mutators from making progress due to lack of memory. Therefore, the SV-GC collector does not guarantee the obstruction-freedom for mutators and must rely heavily on the scheduler to avoid such a scenario.
\footnote{In order to reclaim unreachable cyclic structures of objects, the reference-counting collectors use either a backup tracing collector \cite{AzaLPP03} infrequently or a cycle collector \cite{PazBKPR07}. Both the efficient backup tracing collector \cite{AzaLPP03} and cycle collector \cite{PazBKPR07} use the sliding view technique.}

The basic idea of the sliding view technique in the SV-GC algorithm is as follows. At the beginning of a collection cycle $k$, the collector takes an asynchronous heap snapshot $S_k$ of all (heap) reference slots $s$. By comparing snapshot $S_{k-1}$ and $S_k$, the collector knows which objects have their reference counter changed during the interval between the two collections. For instance, if in the interval a reference slot $s$ is sequentially assigned references to objects $o_0, o_1, \cdots, o_n$, where $(s,o_1)$ is recorded in $S_{k-1}$ and $(s, o_n)$ in $S_k$,  
the collector only needs to execute two reference count updates for $o_0$ and $o_n$: $RC(o_0)--$ and $RC(o_n)++$, instead of $2n$ reference count updates for $o_0$, $o_n$ and $(n-1)$ immediate objects $o_i, 1 \leq i \leq (n-1)$: $RC(o_0)--, RC(o_1)++, RC(o_1)--, \cdots, RC(o_n)++$. The main stages of the {\em generic} sliding view algorithm \cite{LevP06} are shown in Algorithm \ref{alg:GenericCollector}. The algorithm is {\em generic} in the sense that it may use any mechanism for obtaining the sliding view.
Instead of using an atomic snapshot algorithm \cite{AfeADGMS93} to obtain a consistent view of all heap reference slots, the algorithm uses a much simpler mechanism called {\em snooping} \cite{DijLMSS78} to avoid wrong reference counts that result from an inconsistent view. For instance, if the only reference to an  object $O$ is moving from slot $s_1$ to slot $s_2$ when the view is taken, the view may miss the reference in both $s_1$ (reading after modification) and  $s_2$ (reading before modification). To deal with the problem, the snooping mechanism marks as {\em local} any object that is assigned a new reference in the heap while the view is being read from the heap. The marked objects are left to be collected in the next collection cycle. The reader is referred to \cite{LevP06} for the complete SV-GC algorithm.

\begin{algorithm}[t]
  \caption{{\sc GenericCollector}: the main stages of a collection cycle using the sliding view technique} \label{alg:GenericCollector}
  \begin{algorithmic}[1]
	\STATE Raise the $Snoop_i$ flag of each mutator;
	\STATE Obtain a sliding view (concurrently with mutator's computation); \\
	\STATE For each mutator $M_i$: 1) Suspend $M_i$; 2) Turn the $Snoop_i$ flag off; 3) Mark as {\em local} objects $O$ directly reachable from $M_i$'s roots; 4) Resume $M_i$; \\
	\STATE Update the reference counter $O.rc$ of each object $O$; \\
	\STATE Reclaim objects $O$ that are not marked {\em local} and $O.rc=0$; For each descendent $D$ of a reclaimed object, $D.rc --$; $D$ is checked for reclamation like $O$. This operation continues recursively until there are no objects that can be reclaimed. 
  	\end{algorithmic}
\end{algorithm}

We found that the SV-GC algorithm  \cite{LevP06} can be easily improved to provide obstruction-freedom for mutators using the {\em helping technique} \cite{Barnes93}. Basically, if the collector suspends a mutator during its execution of a reference slot update or object allocation procedure, the collector helps the mutator by completing the procedure on behalf of the mutator and moving the mutator's program counter (PC) to the end of the procedure before resuming the mutator. Note that in the concurrent garbage collection model there is only one collector that can suspend a given mutator and the collector suspends only one mutator at a time. The improved algorithm  provides obstruction-freedom for mutators (or application-threads) by preventing mutators from blocking the collector and consequently from blocking other mutators. It is obstruction-free in the sense that progress is guaranteed for each active mutator regardless of the status of the other mutators.


\medskip

{\bf Acknowledgments} Phuong Ha's and Otto Anshus's work was supported by the Norwegian Research Council (grant numbers 159936/V30 and 155550/420). Philippas Tsigas's work was supported by the Swedish Research Council (VR) (grant number 37252706). 

\bibliographystyle{abbrv}
\bibliography{References}

\begin{thebibliography}{10}

\bibitem{AfeADGMS93}
Y.~Afek, H.~Attiya, D.~Dolev, E.~Gafni, M.~Merritt, and N.~Shavit.
\newblock Atomic snapshots of shared memory.
\newblock {\em J. ACM}, 40(4):873--890, 1993.

\bibitem{AKKLYDP93}
A.~Agarwal, J.~Kubiatowicz, D.~Kranz, B.-H. Lim, D.~Yeung, G.~D'Souza, and
  M.~Parkin.
\newblock Sparcle: An evolutionary processor design for large-scale
  multiprocessors.
\newblock {\em IEEE Micro}, 13(3):48--61, 1993.

\bibitem{AlvCCKPS90}
R.~Alverson, D.~Callahan, D.~Cummings, B.~Koblenz, A.~Porterfield, and
  B.~Smith.
\newblock The tera computer system.
\newblock {\em SIGARCH Comput. Archit. News}, 18(3b):1--6, 1990.

\bibitem{Appel04}
A.~W. Appel.
\newblock Real-time concurrent collection on stock multiprocessors.
\newblock {\em SIGPLAN Not.}, 39(4):205--216, 2004.

\bibitem{Asa06}
K.~Asanovic, R.~Bodik, B.~C. Catanzaro, J.~J. Gebis, P.~Husbands, K.~Keutzer,
  D.~A. Patterson, W.~L. Plishker, J.~Shalf, S.~W. Williams, and K.~A. Yelick.
\newblock The landscape of parallel computing research: A view from berkeley.
\newblock {\em Technical Report No. UCB/EECS-2006-183, University of
  California, Berkeley}, 2006.

\bibitem{AttW04}
H.~Attiya and J.~Welch.
\newblock {\em Distributed Computing: Fundamentals, Simulations, and Advanced
  Topics}.
\newblock John Wiley and Sons, Inc., 2004.

\bibitem{AzaLPP03}
H.~Azatchi, Y.~Levanoni, H.~Paz, and E.~Petrank.
\newblock An on-the-fly mark and sweep garbage collector based on sliding
  views.
\newblock In {\em Proc. of the ACM Conf. on Object-oriented Programing,
  Systems, Languages, and Applications (OOPSLA)}, pages 269--281, 2003.

\bibitem{BacALRS01}
D.~F. Bacon, C.~R. Attanasio, H.~B. Lee, V.~T. Rajan, and S.~Smith.
\newblock Java without the coffee breaks: a nonintrusive multiprocessor garbage
  collector.
\newblock In {\em Proc. of the ACM Conf. on Programming Language Design and
  Implementation (PLDI)}, pages 92--103, 2001.

\bibitem{Barnes93}
G.~Barnes.
\newblock A method for implementing lock-free shared-data structures.
\newblock {\em Proc. of the ACM Symp. on Parallel Algorithms and Architectures
  (SPAA)}, pages 261--270, 1993.

\bibitem{BleGV08}
G.~E. Blelloch, P.~B. Gibbons, and S.~H. Vardhan.
\newblock Combinable memory-block transactions.
\newblock In {\em Proc. of the ACM Symp. on Parallel Algorithms and
  Architectures (SPAA)}, pages 23--34, 2008.

\bibitem{BoeDS91}
H.-J. Boehm, A.~J. Demers, and S.~Shenker.
\newblock Mostly parallel garbage collection.
\newblock In {\em Proc. of the ACM Conf. on Programming Language Design and
  Implementation (PLDI)}, pages 157--164, 1991.

\bibitem{CasCSW02}
J.~G. Castanos, L.~Ceze, K.~Strauss, and H.~S.~W. Jr.
\newblock Evaluation of a multithreaded architecture for cellular computing.
\newblock In {\em HPCA '02: Proceedings of the 8th International Symposium on
  High-Performance Computer Architecture}, page 311, 2002.

\bibitem{CheB01}
P.~Cheng and G.~E. Blelloch.
\newblock A parallel, real-time garbage collector.
\newblock In {\em Proc. of the ACM Conf. on Programming Language Design and
  Implementation (PLDI)}, pages 125--136, 2001.

\bibitem{CulSG98}
D.~E. Culler, J.~P. Singh, and A.~Gupta.
\newblock {\em Parallel Computer Architecture: A Hardware/Software Approach}.
\newblock Morgan Kaufmann, 1998.

\bibitem{DalFKLNNDF92}
W.~J. Dally, J.~A.~S. Fiske, J.~S. Keen, R.~A. Lethin, M.~D. Noakes, P.~R.
  Nuth, R.~E. Davison, and G.~A. Fyler.
\newblock The message-driven processor: A multicomputer processing node with
  efficient mechanisms.
\newblock {\em IEEE Micro}, 12(2):23--39, 1992.

\bibitem{DijLMSS78}
E.~W. Dijkstra, L.~Lamport, A.~J. Martin, C.~S. Scholten, and E.~F.~M.
  Steffens.
\newblock On-the-fly garbage collection: an exercise in cooperation.
\newblock {\em Commun. ACM}, 21(11):966--975, 1978.

\bibitem{DolG94}
D.~Doligez and G.~Gonthier.
\newblock Portable, unobtrusive garbage collection for multiprocessor systems.
\newblock In {\em Proc. of the ACM Symp. on Principles of Programming Languages
  (POPL)}, pages 70--83, 1994.

\bibitem{DolL93}
D.~Doligez and X.~Leroy.
\newblock A concurrent, generational garbage collector for a multithreaded
  implementation of ml.
\newblock In {\em Proc. of the ACM Symp. on Principles of Programming Languages
  (POPL)}, pages 113--123, 1993.

\bibitem{DomKP00}
T.~Domani, E.~K. Kolodner, and E.~Petrank.
\newblock A generational on-the-fly garbage collector for java.
\newblock In {\em Proc. of the ACM Conf. on Programming Language Design and
  Implementation (PLDI)}, pages 274--284, 2000.

\bibitem{FeoHKK05}
J.~Feo, D.~Harper, S.~Kahan, and P.~Konecny.
\newblock Eldorado.
\newblock In {\em CF '05: Proceedings of the 2nd conference on Computing
  frontiers}, pages 28--34, 2005.

\bibitem{FraH07}
K.~Fraser and T.~Harris.
\newblock Concurrent programming without locks.
\newblock {\em ACM Trans. Comput. Syst.}, 25(2):5, 2007.

\bibitem{GotGKMRS82}
A.~Gottlieb, R.~Grishman, C.~P. Kruskal, K.~P. McAuliffe, L.~Rudolph, and
  M.~Snir.
\newblock The nyu ultracomputer---designing a mimd, shared-memory parallel
  machine (extended abstract).
\newblock {\em SIGARCH Comput. Archit. News}, 10(3):27--42, 1982.

\bibitem{GscHFHWY06}
M.~Gschwind, H.~Hofstee, B.~Flachs, M.~Hopkins, Y.~Watanabe, and T.~Yamazaki.
\newblock Synergistic processing in cell's multicore architecture.
\newblock {\em Micro, IEEE}, 26(2):10--24, 2006.

\bibitem{GeuHP05}
R.~Guerraoui, M.~Herlihy, and B.~Pochon.
\newblock Polymorphic contention management.
\newblock In {\em Proc. of the Intl. Symp. on Distributed Computing (DISC)},
  pages 303--323, 2005.

\bibitem{GueK08}
R.~Guerraoui and M.~Kapalka.
\newblock On the correctness of transactional memory.
\newblock In {\em Proc. of the ACM Symp. on Principles and Practice of Parallel
  Programming (PPoPP)}, pages 175--184, 2008.

\bibitem{Steele75}
J.~Guy L.~Steele.
\newblock Multiprocessing compactifying garbage collection.
\newblock {\em Commun. ACM}, 18(9), 1975.

\bibitem{HarF03}
T.~Harris and K.~Fraser.
\newblock Language support for lightweight transactions.
\newblock In {\em Proc. of the ACM Conf. on Object-oriented Programing,
  Systems, Languages, and Applications (OOPSLA)}, pages 388--402, 2003.

\bibitem{Her91}
M.~Herlihy.
\newblock Wait-free synchronization.
\newblock {\em ACM Transaction on Programming and Systems}, 11(1):124--149,
  Jan. 1991.

\bibitem{HerLMM05}
M.~Herlihy, V.~Luchangco, P.~Martin, and M.~Moir.
\newblock Nonblocking memory management support for dynamic-sized data
  structures.
\newblock {\em ACM Trans. Comput. Syst.}, 23(2):146--196, 2005.

\bibitem{HerLMS03}
M.~Herlihy, V.~Luchangco, M.~Moir, and I.~William N.~Scherer.
\newblock Software transactional memory for dynamic-sized data structures.
\newblock In {\em Proc. of Symp. on Principles of Distributed Computing
  (PODC)}, pages 92--101, 2003.

\bibitem{KecDMCCL98}
S.~W. Keckler, W.~J. Dally, D.~Maskit, N.~P. Carter, A.~Chang, and W.~S. Lee.
\newblock Exploiting fine-grain thread level parallelism on the mit multi-alu
  processor.
\newblock In {\em Proc. of the Intl. Symp. on Computer Architecture (ISCA)},
  pages 306--317, 1998.

\bibitem{KruRS88}
C.~P. Kruskal, L.~Rudolph, and M.~Snir.
\newblock Efficient synchronization of multiprocessors with shared memory.
\newblock {\em ACM Trans. Program. Lang. Syst.}, 10(4):579--601, 1988.

\bibitem{Lamport76}
L.~Lamport.
\newblock Garbage collection with multiple processes: an exercise in
  parallelism.
\newblock In {\em Proc. of the Intl. Conf. on Parallel Processing}, pages
  50--54, 1976.

\bibitem{Lamport77}
L.~Lamport.
\newblock Concurrent reading and writing.
\newblock {\em Commun. ACM}, 20(11):806--811, 1977.

\bibitem{LevP06}
Y.~Levanoni and E.~Petrank.
\newblock An on-the-fly reference-counting garbage collector for java.
\newblock {\em ACM Trans. Program. Lang. Syst.}, 28(1):1--69, 2006.

\bibitem{MarSS05}
V.~J. Marathe, W.~N.~S. Iii, and M.~L. Scott.
\newblock Adaptive software transactional memory.
\newblock In {\em Proc. of the Intl. Symp. on Distributed Computing (DISC)},
  pages 354--368, 2005.

\bibitem{MicS95}
M.~Michael and M.~Scott.
\newblock Implementation of atomic primitives on distributed shared memory
  multiprocessors.
\newblock In {\em High-Performance Computer Architecture, 1995. Proceedings.,
  First IEEE Symposium on}, pages 222--231, 1995.

\bibitem{Mic04}
M.~M. Michael.
\newblock Hazard pointers: Safe memory reclamation for lock-free objects.
\newblock {\em IEEE Trans. Parallel Distrib. Syst.}, 15(6):491--504, 2004.

\bibitem{CUDA}
NVIDIA.
\newblock {\em NVIDIA CUDA Compute Unified Device Architecture, Programming
  Guide, version 1.1}.
\newblock NVIDIA Corporation, 2007.

\bibitem{PazBKPR07}
H.~Paz, D.~F. Bacon, E.~K. Kolodner, E.~Petrank, and V.~T. Rajan.
\newblock An efficient on-the-fly cycle collection.
\newblock {\em ACM Trans. Program. Lang. Syst.}, 29(4):20, 2007.

\bibitem{PfiBGHKMMNW85}
G.~F. Pfister, W.~C. Brantley, D.~A. George, S.~L. Harvey, W.~J. Kleinfelder,
  K.~P. McAuliffe, E.~S. Melton, V.~A. Norton, and J.~Weiss.
\newblock The ibm research parallel processor prototype (rp3): Introduction and
  architecture.
\newblock In {\em ICPP}, pages 764--771, 1985.

\bibitem{Plo89}
S.~A. Plotkin.
\newblock Sticky bits and universality of consensus.
\newblock In {\em Proc. of Symp. on Principles of Distributed Computing
  (PODC)}, pages 159--175, 1989.

\bibitem{RieFF06}
T.~Riegel, P.~Felber, and C.~Fetzer.
\newblock A lazy snapshot algorithm with eager validation.
\newblock In {\em Proc. of the Intl. Symp. on Distributed Computing (DISC)},
  pages 284--298, 2006.

\bibitem{SinBWC07}
J.~Singer, G.~Brown, I.~Watson, and J.~Cavazos.
\newblock Intelligent selection of application-specific garbage collectors.
\newblock In {\em Proc. of the Intl. Symp. on Memory management (ISMM)}, pages
  91--102, 2007.

\bibitem{Smi85}
B.~Smith.
\newblock The architecture of hep.
\newblock In {\em Parallel MIMD Computation: HEP Supercomputer and Its
  Applications, Scientific Computation Series}, page 41–55, 1985.

\bibitem{SomDK06}
S.~Soman, L.~Dayn\`{e}s, and C.~Krintz.
\newblock Task-aware garbage collection in a multi-tasking virtual machine.
\newblock In {\em ISMM '06: Proc. of the Intl. Symp.on Memory Management},
  pages 64--73, 2006.

\bibitem{SomK07}
S.~Soman and C.~Krintz.
\newblock Application-specific garbage collection.
\newblock {\em J. Syst. Softw.}, 80(7):1037--1056, 2007.

\bibitem{SriRK07}
S.~Sridharan, A.~Rodrigues, and P.~Kogge.
\newblock Evaluating synchronization techniques for light-weight
  multithreaded/multicore architectures.
\newblock In {\em Proc. of the ACM Symp. on Parallel Algorithms and
  Architectures (SPAA)}, pages 57--58, 2007.

\bibitem{CSX06}
C.~Technology.
\newblock Csx processor architecture whitepaper.
\newblock 2006.

\bibitem{SchS05}
I.~William N.~Scherer and M.~L. Scott.
\newblock Advanced contention management for dynamic software transactional
  memory.
\newblock In {\em Proc. of Symp. on Principles of Distributed Computing
  (PODC)}, pages 240--248, 2005.

\bibitem{ZhuSHG07}
W.~Zhu, V.~C. Sreedhar, Z.~Hu, and G.~R. Gao.
\newblock Synchronization state buffer: supporting efficient fine-grain
  synchronization on many-core architectures.
\newblock In {\em Proc. of the Intl. Symp. on Computer Architecture (ISCA)},
  pages 35--45, 2007.

\end{thebibliography}

\end{document}